\documentclass[10pt,conference,twocolumn]{IEEEtran}
\usepackage{filecontents}
\usepackage[noadjust]{cite}
\usepackage{caption}
\usepackage{amsfonts}
\usepackage{subcaption}
\usepackage[dvips]{color, graphicx}
\usepackage[cmex10]{amsmath}
\usepackage{pgfplotstable}
\usepackage{array}
\usepackage{booktabs}
\usepackage{times}
\usepackage[normalem]{ulem} 
\usepackage[lined,boxed,commentsnumbered]{algorithm2e}
\usepackage{flushend}

\long\def\symbolfootnote[#1]#2{\begingroup%
\def\thefootnote{\fnsymbol{footnote}}\footnote[#1]{#2}\endgroup}

\begin{document}
\title{A Novel Interleaving Scheme for Polar Codes}
\author{\IEEEauthorblockN{Ya Meng\IEEEauthorrefmark{1},
Liping Li\IEEEauthorrefmark{1},
Yanjun Hu\IEEEauthorrefmark{1}}
\IEEEauthorblockA{\IEEEauthorrefmark{1}
Key Laboratory of Intelligent Computing and Signal Processing of the Ministry\\
of Education of China Anhui University, China,\\
Email: mengya@ahu.edu.cn, liping\_li@ahu.edu.cn,  yanjunhu@ahu.edu.cn}}

\maketitle
\begin{abstract}
It's known that the bit errors of polar codes with successive cancellation (SC)
decoding are coupled. We call the coupled
information bits the correlated bits. In this paper, concatenation schemes are studied
for polar codes (as inner codes) and LDPC codes (as outer codes).
In a conventional concatenation scheme, to achieve a better BER performance,
one can divide all $N_l$ bits in a LDPC block into $N_l$ polar blocks to
completely de-correlate the possible coupled errors.
In this paper, we propose a novel interleaving scheme between a LDPC code  and
a polar code which breaks the correlation of the errors among the
correlated bits. This interleaving scheme still keeps the simple SC decoding of polar codes while
achieves a comparable BER performance at a much smaller delay compared with a $N_l$-block delay scheme.
\end{abstract}

\section{Introduction}\label{sec_ref}
Polar codes are proposed by Ar$\i$kan in \cite{arikan_iti09} which provably achieve the capacity of symmetric binary-input discrete memoryless channels (B-DMCs) with a low encoding and decoding complexity. The encoding and decoding process (with successive cancellation, SC) can be implemented with a complexity of $\mathcal{O}(N \log N)$. The idea of polar codes is to transmit information bits on those noiseless channels while fixing the information bits on those completely noisy channels. The fixed bits are made known to both the transmitter and receiver.

To improve the polar code performance in the finite domain, various decoding
processes \cite{arikan_icl08,urbanke_isit09,vardy_it15,niu_itc13}
and concatenation schemes \cite{eslami_isit11,guo_isit14,barry_icc13} were proposed. The decoding processes in
these works have higher complexity than the original SC decoding of \cite{arikan_iti09}.
Systematic polar codes are later proposed in \cite{arikan_icl11} which have almost the same decoding complexity.
In non-systematic encoding, the codeword $\mathbf{x}$ is obtained by $\mathbf{x}=\mathbf{u}G$, where $G$ is the generator matrix. The basic idea of systematic polar codes is to use some part of the codeword $\mathbf{x}$ to transmit information bits instead of directly using the source bits $\mathbf{u}$ to transmit them. In \cite{arikan_icl11}, it's shown that systematic polar codes achieve better BER performance than non-systematic polar codes. But, theoretically, this better BER performance is not expected from the indirect decoding process: first decoding $\hat{\mathbf{u}}$ ($\hat{\mathbf{u}}$ is the estimation
of $\mathbf{u}$ from the normal SC decoding) then re-encoding $\hat{\mathbf{x}}$ as $\hat{\mathbf{u}}G$. One would expect that any errors in $\hat{\mathbf{u}}$ would be amplified in this re-encoding process.

In \cite{li_jsac15}, the reason that systematic polar codes have better BER performance than non-systematic polar codes has been studied. From \cite{li_jsac15}, we already see that the number of errors of polar codes with the SC decoding is not necessarily amplified in the indirect decoding process of systematic polar codes. It all depends on how the errors are distributed in the SC decoding process. From the re-encoding $\hat{\mathbf{x}}=\hat{\mathbf{u}}G$ and
that the number of errors in $\hat{\mathbf{x}}$ is smaller than that of $\hat{\mathbf{u}}$, we can conclude that the
coupling of the errors in $\hat{\mathbf{u}}$ are controlled by the columns of $G$. We provide a proposition of this
coupling pattern in this paper. Based on this coupling pattern, a novel interleaving scheme is introduced to
improve the performance of polar codes with finite block lengths while still maintaining the low complexity of the SC
decoding. We use a LDPC code as the outer code and a polar code as the inner code. Note that the
concatenation of polar codes with LDPC codes is studied in \cite{eslami_isit11} and \cite{guo_isit14} where
no interleaving is used and BP (belief-propagation) decoding is applied for polar codes.

The advantage of the proposed interleaving scheme between LDPC codes and polar codes is that
the coupled errors between the correlated information bits are divided into different LDPC blocks. Therefore,
to achieve the same BER performance as a conventional concatenation scheme, either a simpler LDPC code
or a larger code rate can be applied. The proposed interleaving scheme achieves a comparable BER performance
as a blind interleaving scheme where all the $K$ information bits of polar codes are de-coupled into $K$ LDPC
blocks. We provide simulation results to verify our interleaving scheme.

Following the notations in \cite{arikan_iti09}, in this paper,
we use $v_1^N$ to represent a row vector with elements $(v_1,v_2,...,v_N)$. We also use
$\mathbf{v}$ to represent the same vector for notational convenience. Suppose a vector $v_1^N$, the
vector $v_i^j$ is a subvector $(v_i, ..., v_j)$ with $1 \le i,j \le N$. If there is a set $\mathcal{A} \in \{1,2,...,N\}$,
then $v_{\mathcal{A}}$ denotes a subvector with elements in $\{v_i, i \in \mathcal{A}\}$.

The rest of the paper is organized as follows. Section \ref{sec_background} introduces the fundamentals of
non-systematic and systematic polar codes.  Also introduced in this section is the coupling pattern of
polar codes with the SC decoding.  Section \ref{sec_interleaving} proposes the new interleaving scheme
with a detailed algorithm.
Section \ref{sec_result} presents the simulation results. Finally the conclusion remarks are provided at the end.

\section{Systematic Polar Codes}\label{sec_background}

In the first part of this section, the relevant theories on non-systematic polar codes and systematic polar codes are presented.
In the second part of this section, the correlation among errors is introduced, which is the basis of the proposed interleaving scheme.

\subsection{Preliminaries of Non-Systematic Polar Codes}\label{sec_background_nonsys}
The generator matrix for polar codes is $G_N= BF^{\otimes n}$ where $B$ is a bit-reversal
matrix, $F=\bigl(\begin{smallmatrix} 1&0 \\ 1&1\end{smallmatrix}\bigr)$, $n=\log_2N$, and
$F^{\otimes n}$ is the $n$th Kronecker power of the matrix $F$ over the binary field $\mathbb{F}_2$.
In this paper, we consider an encoding matrix $G=F^{\otimes n}$ without the permutation matrix $B$. This matrix is the basis of analyzing the interleaving scheme of this paper.

Mathematically, the encoding is a process to obtain the codeword $\mathbf{x}$ through $\mathbf{x} = \mathbf{u}G$
for a given source vector $\mathbf{u}$. The source vector $\mathbf{u}$
consists of the information bits and the frozen bits,
denoted by ${u}_{\mathcal{A}}$ and ${u}_{\bar{\mathcal{A}}}$,
respectively. Here the set $\mathcal{A}$ includes the indices for the information bits and
$\bar{\mathcal{A}}$ is the complementary set.
The set $\mathcal{A}$ can be constructed by
selecting indices of the bit channels with the smallest Bhattacharyya parameters. In \cite{arikan_iti09}, there are detailed definition of the bit channels and the corresponding Bhattacharyya parameters.
 Both sets $\mathcal{A}$ and $\bar{\mathcal{A}}$  are in $\{1, 2,..., N\}$
for polar codes with a block length $N=2^n$.

\begin{figure}
{\par\centering
\resizebox*{3.0in}{!}{\includegraphics{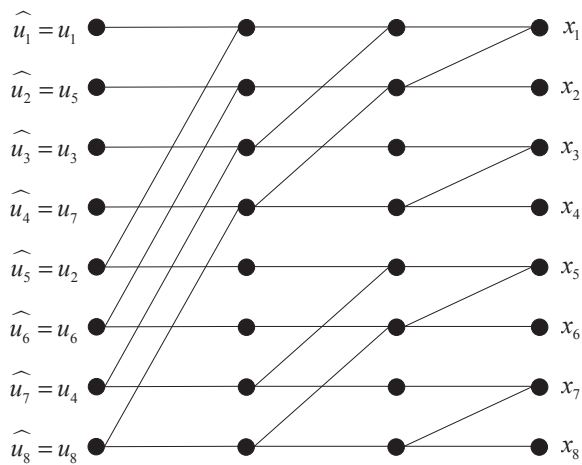}} \par}
\caption{An encoding circuit of the non-systematic polar codes with $N=8$. Signals flow
from the left to the right. Each edge carries a signal of 0 or 1.}
\label{fig_encoding_nonsys}
\end{figure}

An encoding circuit is shown in Fig.~\ref{fig_encoding_nonsys}.
If the nodes in Fig.~\ref{fig_encoding_nonsys} are viewed as memory elements, the encoding
process is to calculate the corresponding values to fill all the memory elements from the left to the right.

\subsection{Construction of Systematic Polar Codes}\label{sec_background_sys}
For systematic polar codes,
we also focus on a generator matrix without the permutation matrix $B$, namely $G=F^{\otimes n}$.

The source bits $\mathbf{u}$ can be
split as $\mathbf{u} = ({u}_{\mathcal{A}}, {u}_{\bar{\mathcal{A}}})$. The first part ${u}_{\mathcal{A}}$ consists of user data
that are free to change in each round of transmission, while the
second part ${u}_{\bar{\mathcal{A}}}$  consists of data that are
frozen at the beginning of each session and made known to the decoder.
The codeword can then be expressed as
\begin{equation} \label{eq_x2}
\mathbf{x} = {u}_{\mathcal{A}}G_{\mathcal{A}}+ {u}_{\bar{\mathcal{A}}}G_{\bar{\mathcal{A}}}
\end{equation}
where $G_{\mathcal{A}}$ is the sub-matrix of $G$ with rows specified by the set $\mathcal{A}$.
The systematic polar code is constructed by
specifying a set of indices of the codeword $\mathbf{x}$ as the indices to convey the information bits. Denote this
set as $\mathcal{B}$ and the complementary set as $\bar{\mathcal{B}}$. The codeword $\mathbf{x}$ is thus split as
$({x}_{\mathcal{B}},{x}_{\bar{\mathcal{B}}})$. With some manipulations, we have
\begin{flalign}
\begin{split} \label{eq_xb_xbc}
{x}_{\mathcal{B}}={u}_{\mathcal{A}}G_{\mathcal{AB}}+{u}_{\bar{\mathcal{A}}}G_{\bar{\mathcal{A}}\mathcal{B}}\\ {x}_{\bar{\mathcal{B}}}={u}_{\mathcal{A}}G_{\mathcal{A\bar{B}}}+{u}_{\bar{\mathcal{A}}}G_{\mathcal{\bar{A}\bar{B}}}
\end{split}
\end{flalign}
The matrix $G_{\mathcal{AB}}$ is a sub-matrix of the generator matrix with elements
$\{G_{i,j}\}_{i \in \mathcal{A}, j \in \mathcal{B}}$. Given a non-systematic encoder $(\mathcal{A},u_{\mathcal{\bar{A}}})$,
there is a systematic encoder $(\mathcal{B},u_{\mathcal{\bar{A}}})$ which performs the mapping
${x}_{\mathcal{B}} \mapsto {\mathbf{x}}=({x}_{\mathcal{B}},{x}_{\bar{\mathcal{B}}})$. To realize this systematic
mapping, ${x}_{\bar{\mathcal{B}}}$ needs to be computed for any given information bits ${x}_{\mathcal{B}}$. To this
end, we see from (\ref{eq_xb_xbc}) that ${x}_{\bar{\mathcal{B}}}$ can be computed if $u_{\mathcal{A}}$ is known.
The vector $u_{\mathcal{A}}$ can be obtained as the following
\begin{equation} \label{eq_ua}
u_{\mathcal{A}} = (x_{\mathcal{B}}-u_{\bar{\mathcal{A}}}G_{\mathcal{\bar{A}B}})(G_{\mathcal{AB}})^{-1}
\end{equation}
From (\ref{eq_ua}), it's seen that $x_{\mathcal{B}} \mapsto u_{\mathcal{A}}$ is one-to-one if $x_{\mathcal{B}}$
has the same elements as $u_{\mathcal{A}}$ and if $G_{\mathcal{AB}}$ is invertible.
In \cite{arikan_icl11}, it's shown that $\mathcal{B} = \mathcal{A}$ satisfies all these conditions in order to establish the
one-to-one mapping $x_{\mathcal{B}} \mapsto u_{\mathcal{A}}$. In the rest of the paper, the systematic encoding of polar
codes adopts this selection of $\mathcal{B}$ to be $\mathcal{B} = \mathcal{A}$. Therefore we can rewrite (\ref{eq_xb_xbc}) as
\begin{flalign}
\begin{split} \label{eq_xb_xbc_2}
{x}_{\mathcal{A}} ={u}_{\mathcal{A}}G_{\mathcal{AA}}+{u}_{\bar{\mathcal{A}}}G_{\bar{\mathcal{A}}\mathcal{A}}\\ {x}_{\bar{\mathcal{A}}} ={u}_{\mathcal{A}}G_{\mathcal{A\bar{A}}}+{u}_{\bar{\mathcal{A}}}G_{\mathcal{\bar{A}\bar{A}}}
\end{split}
\end{flalign}

\subsection{Correlated Bits}\label{sec_gene_matrix}
In \cite{arikan_icl11}\cite{li_jsac15}, it's shown that the re-encoding process
of $\hat{\mathbf{x}}=\hat{\mathbf{u}}G$ after decoding $\hat{\mathbf{u}}$ does not amplify the number of
errors in $\hat{\mathbf{u}}$.
Instead, there are less errors in $\hat{\mathbf{x}}$ than in $\hat{\mathbf{u}}$. This clearly
shows that the coupled errors in $\hat{\mathbf{u}}$ are de-coupled (or cancelled) in the re-encoding
process. In this section, we first restate a corollary from \cite{li_jsac15} and then provide a proposition to
show the coupling pattern of the errors in $\hat{\mathbf{u}}$. This coupling pattern is used in Section
\ref{sec_interleaving} to design the interleaving scheme.
\newtheorem{corollary}{Corollary}
\begin{corollary}
The matrix $G_{\mathcal{\bar{A}}\mathcal{A}} = 0$.
\end{corollary}
The proof of this corollary can be found in \cite{li_jsac15}. The following proposition shows the pattern of the
coupling of the errors in $\hat{\mathbf{u}}$ from the SC decoding process.
\newtheorem{proposition}{Proposition}
\begin{proposition}\label{proposition_1}
Let the indices of the non-zero entries of column $i \in \mathcal{A}$ of $G_N$ be $\mathcal{A}_i$.
Then, the errors of $\hat{{u}}_{\mathcal{A}_i}$ are dependent.
\end{proposition}

The proof of Proposition \ref{proposition_1} is omitted in the current paper due to the space limit.
From Proposition \ref{proposition_1}, an error pattern among the errors in $\hat{\mathbf{u}}$ is
shown. We call bits $\hat{{u}}_{\mathcal{A}_i}$ the correlated estimated bits. This says that statistically,
the errors of bits $\hat{{u}}_{\mathcal{A}_i}$ are coupled. To show this coupling, we give an
example of $N=16$ and $R=0.5$ in a BEC channel with an erasure probability $0.2$. The indices selected
in this case for information bits are $\{8,10,11,12,13,14,15,16\}$ (indexed from 1 to 16). The
coupling effect (similar to the correlation coefficient) of bits indicated by non-zero positions
of column 10, 11, and 13
is recorded in simulations and is shown in Table \ref{label_relevant}. From Table \ref{label_relevant}, the coupling of the errors in column 10, 11, and 13 is clearly shown.

\begin{table}
\centering
\caption{Coupling effect for $N=16$, $R=0.5$ in a BEC channel with an erasure probability of 0.2}
\label{label_relevant}
\begin{tabular}{|c|c|}
\hline
 columns of G  &  coupling coefficient \\
\hline
 10 & 76\%  \\
\hline
 11 & 74\%  \\
\hline
 13 & 74\%  \\
\hline
\end{tabular}
\end{table}

To the authors' knowledge, there is no
attempt yet to utilize this coupling pattern to improve the performance of polar codes. In the next section
of this paper, we propose a novel interleaving scheme to break the coupling of errors to improve the BER
performance of polar codes while still maintaining the low complexity of the SC decoding.

%

\section{The Proposed Interleaving Scheme}\label{sec_interleaving}
In this section we consider an interleaving scheme between a LDPC code (the outer code) and
a polar code (the inner code). From Proposition \ref{proposition_1}, we know
the exact correlated information bits of the polar codes. The task of the interleaving scheme is
thus to make sure that the correlated bits of the inner polar codes come from different
LDPC blocks. In this way, the de-interleaved LDPC blocks have independent errors.
We call this interleaving scheme the correlation-breaking interleaving (CBI).
Before explaining our interleaving scheme, a blind interleaving (BI) is introduced which
breaks all  bits in one LDPC block into different polar code blocks. The scheme guarantees that
errors in each LDPC block are independent. This BI scheme serves as a benchmark for our CBI scheme.

\subsection{The Blind Interleaving Scheme}\label{sec_all_scheme}
\begin{figure}
{\par\centering
\resizebox*{3.0in}{!}{\includegraphics{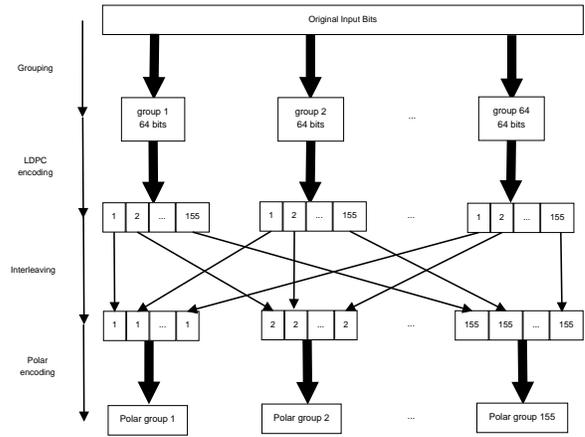}} \par}
\caption{
An blind interleaving (BI) scheme. The block length of the LDPC code is $N_l=155$, and the code rate is $64/155$.
The block length of the polar code is $N=256$, and the code rate is $R=1/4$.
}
\label{fig_all_bits_permutation}
\end{figure}

In this section, the scheme of scattering all bits in a LDPC block into different polar code blocks is introduced.
Suppose one LDPC
code block has $K_l$ information bits and the block length is $N_l$. These $N_l$ bits are divided into $N_l$ polar code
blocks, which guarantees that the errors in each LDPC block are independent as they come from different polar code blocks.
We give an example in Fig.~\ref{fig_all_bits_permutation} where $K_l=64$ and $N_l=155$.
In Fig.~\ref{fig_all_bits_permutation}, bit $i$ of all the $K_l=64$ LDPC code blocks form the input vector
to the $i$th polar code encoder. Polar code in this example has $N=256$ and a rate $R=1/4$.
In the receiver side, after the polar codes decoding, a de-interleaving is done to collect the $N_l = 155$ outputs
for one LDPC block from $N_l=155$ polar code blocks. For one LDPC block, the delay is thus $N_l=155$ polar code blocks.


\subsection{The Correlation-Breaking Interleaving (CBI) Scheme}\label{sec_part_scheme}

The BI scheme in Section \ref{sec_all_scheme} has a long delay. From Section \ref{sec_gene_matrix}, we
know that it is not necessary to scatter all bits in a LDPC block into different polar blocks since not all bits in a polar
block are correlated. Only those bits in $\{\mathcal{A}_i\}_{i=1}^K$ are correlated.
The interleaving scheme in this section is
to make the bits $\{\mathcal{A}_i\}_{i=1}^K$ of one polar block composed of
different LDPC blocks. Or in other words, the interleaving
scheme is to scatter the information bits $\{\mathcal{A}_i\}_{i=1}^K$ of each polar block into different LDPC blocks.

The difficulty in designing a CBI scheme is that the sets $\{\mathcal{A}_i\}_{i=1}^K$ are different
for different block lengths and data rates. They are also different for different underlying channels for
which polar codes are designed.  A CBI scheme is dependent on at least three parameters:
the block length $N$, the data rate $R$, and the underlying channel $W$.
Let's denote a CBI scheme as CBI($N$,$R$,$W$) to show this dependence.
A CBI($N$,$R$,$W$) optimized for one
set of ($N$,$R$,$W$) is not
necessarily optimized for  another set ($N'$,$R'$,$W'$). It may not even work for the set ($N'$,$R'$,$W'$)
if $N'R' \neq NR$. In the following, we provide a CBI scheme which works for any sets of ($N$,$R$,$W$), but
not necessarily optimal for one specific set of ($N$,$R$,$W$).

As $\mathcal{A}_i$ are the indices of the non-zero entries of column $i \in \mathcal{A}$, we first
extract the $K=|\mathcal{A}|$ columns of $G$ and denote it as the submatrix $G(:,\mathcal{A})$.
Divide the submatrix $G(:,\mathcal{A})$ as $G(:,\mathcal{A}) = [G_{\mathcal{\bar{A}A}} ~~ G_{\mathcal{AA}}]$.
Since the submatrix $G_{\mathcal{\bar{A}A}} = \mathbf{0}$, we only need to analyze the
submatrix $G_{\mathcal{AA}}$. If a CBI needs to look at each individual set $\mathcal{A}_i$, then
a general CBI is beyond reach. However, we can simplify this problem by dividing
the information bits only into two groups: the correlated bits $\mathcal{A}_c$ and the uncorrelated bits $\bar{\mathcal{A}_c}$.
The following proposition can be used to find the sets $\mathcal{A}_c$ and $\bar{\mathcal{A}_c}$.
\begin{proposition}\label{proposition_2}
For the submatrix $G_{\mathcal{AA}}$, the row indices (relative to the submatrix $G_{\mathcal{AA}}$) with Hamming weight greater than one is denoted as the set $\tilde{\mathcal{A}_c}$. The corresponding set of $\tilde{\mathcal{A}_c}$ with respect to the matrix $G$ is the set $\mathcal{A}_c$.
\end{proposition}

The proof of Proposition \ref{proposition_2} is omitted in this paper due to the space limit.
We give an example of how to use Proposition \ref{proposition_2} to find the set $\mathcal{A}_c$ and $\bar{\mathcal{A}_c}$.
Let the block length be $N=16$, the code rate $R=0.5$, and the underlying channel is the BEC channel with an
erasure probability 0.2. The set $\mathcal{A}$ is the same as the example in Section \ref{sec_gene_matrix}.
With Proposition \ref{proposition_2}, we can easily find that $\tilde{\mathcal{A}_c}=\{4,6,7,8\}$ for
the submatrix $G_{\mathcal{AA}}$. Relative to the matrix $G_{16}$, this set is $\mathcal{A}_c=\{12,14,15,16\}$.
The uncorrelated set is thus $\bar{\mathcal{A}_{c}}=\{8,10,11,13\}$.

\begin{figure}
{\par\centering
\resizebox*{3.0in}{!}{\includegraphics{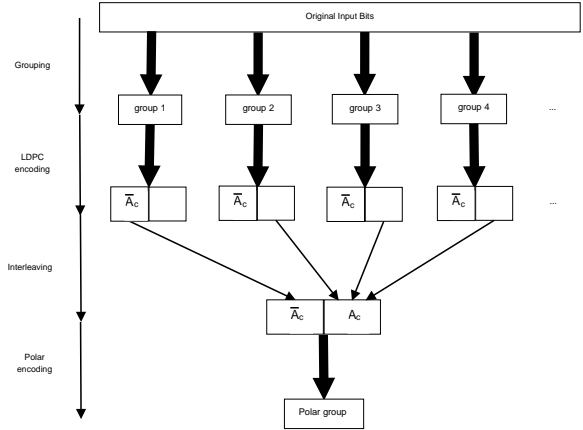}} \par}
\caption{
A general correlation-breaking interleaving scheme.
Here the set $\mathcal{A}_c$ consists of the indices of the correlated bits and the set
$\bar{\mathcal{A}_c}$ is the complementary set of $\mathcal{A}_c$.
}
\label{fig_correlated}
\end{figure}

With the sets $\mathcal{A}_c$ and $\bar{\mathcal{A}_c}$ obtained for any ($N$,$R$,$W$), we can devise a CBI scheme.
Let $K_c = |\mathcal{A}_c|$ and $K_{uc} = |\bar{\mathcal{A}_c}|$. Fig.~\ref{fig_correlated} is a general CBI scheme.
And Algorithm \ref{algorithm_CBI} shows a detailed implementation. Algorithm \ref{algorithm_CBI} contains two parts.
Part one (from line 5 to 16) is collecting the encoded LDPC bits as the uncorrelated information bits of polar groups.
Part two (from line 17 to the end) collects the encoded LDPC bits for the correlated information bits of polar groups.
The principle is of course that: the correlated information bits of polar codes come from different LDPC blocks while
the uncorrelated information bits can be from the same LDPC block. Note that the $K_{uc}$ uncorrelated information bits
of one polar block can be directly taken from a continuous chunk of a LDPC block. However, taking the $K_c$ correlated
information bits for each polar encoding block needs a fine design. In Algorithm \ref{algorithm_CBI}, two new
sets $\mathcal{A}_{cc}$ (line 19) and $\mathcal{A}_{cp}$ (line 22) are defined which control the collecting
of the correlated information bits for each polar encoding block.

The CBI scheme in Algorithm \ref{algorithm_CBI} runs every $K_n=K_c+1$ LDPC blocks. For each of these LDPC blocks,
$n_d*K_n+mo$ polar blocks are needed where $n_d$, $K_n$ and $mo$ are defined at the top of
Algorithm \ref{algorithm_CBI}. The average delay looking from the LDPC side (in terms of the polar
blocks) is: $\lceil(n_d*K_n+mo)/K_n\rceil \le n_d+1$.

\IncMargin{1em}
\LinesNumbered
\begin{algorithm}
\SetKwInOut{Input}{INPUT} \SetKwInOut{Output}{OUTPUT}
\Input{$N_l$, $K$, $K_c$, $K_{uc}$, $\mathcal{A}_{c}$,} 
\Output{

the input bits $\mathbf{U}_{\mathcal{A}}$ of $\lfloor N_l/K \rfloor*(K_c+1)+N_l\%K$\
groups of polar codes}
\BlankLine
$n_d$=$\lfloor N_l/K \rfloor$ \;
$n_u$=$\lceil N_l/K \rceil$ \;
$mo$=$N_l\%K$ (\tcp*[h]{take $N_l$ modulo $K$})\;
$K_n$=$K_c+1$\;
\For (\tcp*[h]{collect bits from LDPC blocks as input bits $\bar{\mathcal{A}_{c}}$ for the $ith$ polar encoding}) {$j \leftarrow 1$ \KwTo $n_d$}{
        \For {$i \leftarrow (j-1)*K_n+1$ \KwTo $(j-1)*K_n+K_n$}{
                 $ls=((i-1)\%K_n)*N_l$\;
                 $\mathbf{U}_{\mathcal{A}}((i-1)*K+\bar{\mathcal{A}_{c}})$=$U_{L}(ls+i-j+K_{uc}*(j-1)+1:ls+i-j+K_{uc}*(j-1)+1+K_{uc}-1)$\;
            }
       }
\If {$mo!=0~\&\&~j==n_u$}{
\For {$i \leftarrow (n_u-1)*K_n+1$ \KwTo $(n_u-1)*K_n+mo$}{
    $ls=((i-1)\%K_n)*N_l$\;
    $\mathbf{U}_{\mathcal{A}}((i-1)*K+\bar{\mathcal{A}_{c}})$= $U_{L}(ls+i-n_u+K_{uc}*(n_u-1)+1:ls+i-n_u+K_{uc}*(n_u-1)+1+(K_{uc}-i\%K_n-1))$\;
}
}

\For (\tcp*[h]{collect bits from LDPC blocks as input bits $\mathcal{A}_{c}$ for the $ith$ polar encoding}){$j \leftarrow 1$ \KwTo $n_d$}{
        \For {$i \leftarrow (j-1)*K_n+1$ \KwTo $(j-1)*K_n+K_n$}{
                $\mathcal{A}_{cc}$=$\{1,2,...,i-K_n*(j-1)-1\}$\;
                $lc=(\mathcal{A}_{cc}-1)*N_l$\;
                $\mathbf{U}_{\mathcal{A}}((i-1)*K+\mathcal{A}_{c}(\mathcal{A}_{cc}))$=$U_{L}(lc+i-j+K_{uc}*j)$\;
                $\mathcal{A}_{cp}$=$\{(i-K_n*(j-1)+1),...,K_n\}$\;
                $lm=(\mathcal{A}_{cp}-1)*N_l$\;
                $\mathbf{U}_{\mathcal{A}}((i-1)*K+\mathcal{A}_{c}(\mathcal{A}_{cp}-1))$=$U_{L}(lm+i-(j-1)+K_{uc}*(j-1))$\;
                }
            }
\If {$mo!=0~\&\&~j==n_u$}{
     \For {$i \leftarrow (n_u-1)*K_n+1$ \KwTo $(n_u-1)*K_n+mo$}{
                $\mathcal{A}_{cp}$=$\{(i-K_n*(n_u-1)+1),...,K_n\}$\;
                $lm=(\mathcal{A}_{cp}-1)*N_l$\;
                $\mathbf{U}_{\mathcal{A}}((i-1)*K+\mathcal{A}_{c}(\mathcal{A}_{cp}-1))$=$U_{L}(lm+i-(n_u-1)+K_{uc}*(n_u-1))$\;
                    }
}
\caption{ 
The algorithm of a general correlation-breaking interleaving scheme. Transmitting $K_n$  LDPC blocks needs
$n_d*K_n+mo$ groups of polar codes where $K_n$, $n_d$, and $mo$ are defined at the top of this algorithm.
}
\label{algorithm_CBI}
\end{algorithm} \DecMargin{1em}

\section{Simulation Result}\label{sec_result}
In this section, simulation results are provided to verify the performance of the CBI
scheme shown in Algorithm \ref{algorithm_CBI}. The example we take is the
same as the BI scheme in Fig.~\ref{fig_all_bits_permutation}. All the LDPC codes used
in this section is the (155,64,20)
Tanner code \cite{tanner_it04}.  Therefore $N_l=155$ and $K_l=64$.
The polar code has the $N=256$ and $R=1/4$. The underlying channel is the AWGN channel.
The polar code construction is based on \cite{vardy_it13} which produces the set $\mathcal{A}$.
Then the submatrix $G_{\mathcal{AA}}$ is formed from the generator matrix $G_{256}$. Based
on the submatrix   $G_{\mathcal{AA}}$,
the correlated set $\mathcal{A}_c$ ($K_c = 36$) and the un-correlated set $\bar{\mathcal{A}_c}$ ($K_{uc} = 28$) is obtained.
Algorithm \ref{algorithm_CBI} is implemented with the following details.

\begin{itemize}
\item Consider the $i$th polar encoding block for $1 \le i \le K_c + 1 = 37$.
The information bits $\bar{\mathcal{A}_{c}}$ of the $i$th polar block
is composed of bit $i$ to $(i+28-1)$ of the $((i-1)\%37+1)$th LDPC code block.
The information bits $\mathcal{A}_{c}$ for the $i$th polar block are collected through two sets $\mathcal{A}_{cc}$ and
$\mathcal{A}_{cp}$ with $\mathcal{A}_{cc}=\{i-1,i-2,...1\}$ and $\mathcal{A}_{cp}=\{37,36,35...i+1\}$.
These two sets $\mathcal{A}_{cc}$ and $\mathcal{A}_{cp}$ are the indices of LDPC blocks.
The bits of $\mathcal{A}_c$ of the $ith$ polar block are from two parts:
the $(i-1+28)$th bit of LDPC groups $\mathcal{A}_{cc}$
and the $i$th bit of LDPC groups $\mathcal{A}_{cp}$.

\item Consider the $i$th polar group for $38 \le i \le 74$. The information bits  $\bar{\mathcal{A}_{c}}$ for
the $ith$ polar code consists of bits $(i-2+28+1)$ to $((i-2+28+1)+28-1)$ of the $((i-1)\%37+1)$th LDPC block.
In this case $\mathcal{A}_{cc}=\{(i-37)-1,(i-37)-2,...1\}$ and $\mathcal{A}_{cp}=\{37,36,35,...(i-37)+1\}$.
Therefore the bits $\mathcal{A}_c$ of the $i$th polar code are from bit $(i-2+28*2)$ of LDPC groups
$\mathcal{A}_{cc}$ and bit $(i-1+28)$ of LDPC groups $\mathcal{A}_{cp}$.

\item Now consider the $i$th polar group for  $75 \le i \le 101$.
 The bits $\bar{\mathcal{A}_{c}}$ of the $ith$ polar code is made up of bits from
  $(i-3+28*2+1)$ to $((i-3+28*2+1)+(28-i\%37)-1)$ of the $((i-1)\%37+1)$th LDPC block.
   In this case, there is only $\mathcal{A}_{cp}=\{37,36,35,...(i-37*2)+1\}$.
   The information bits $\mathcal{A}_c$ for the $ith$ polar code are from bit $(i-2+28*2)$
   of LDPC groups $\mathcal{A}_{cp}$.
\end{itemize}

\begin{figure}
{\par\centering
\resizebox*{3.0in}{!}{\includegraphics[clip]{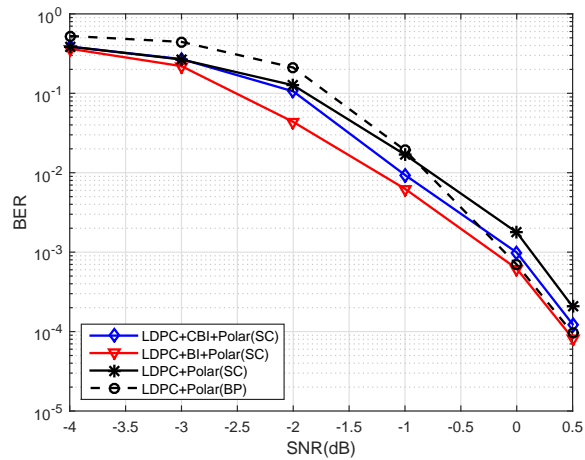}} \par}
\caption{The BER performance of a polar code concatenated with a LDPC code in AWGN channels. The polar code
has a block length $N=256$ and the data rate $R=1/4$. The LDPC code is the (155,64,20) Tanner code.}
\label{fig_simulation_result}
\end{figure}

In this example, the average delay in terms of the polar blocks is $n_d+1 = 3$, which is much smaller than $N_l=155$.
The BER performance of the CBI scheme is shown in Fig.~\ref{fig_simulation_result} where the solid line
with diamonds is the performance of the CBI scheme. The legend for this scheme is: LDPC+CBI+Polar(SC).
The solid line with stars is the
the performance of polar code directly concatenated with the LDPC code (no interleaving
being performed), with a legend of LDPC+Polar(SC).
Note that compared with the BI scheme (the solid line with
triangles and the legend LDPC+BI+Polar(SC)),
the CBI scheme achieves almost the same performance while having a delay $N_l/(n_d+1) = 51$ times
smaller. Both the CBI and the BI scheme has a better BER performance than the direct concatenation.

To compare with a direct
concatenation of polar codes (BP decoding) with LDPC codes, the same simulation is carried out
(shown in Fig.~\ref{fig_simulation_result} by the dashed circled line).
The legend for this scheme is LDPC+Polar(BP).
At low SNR regions, the LDPC+CBI+Polar(SC) scheme outperforms the LDPC+Polar(BP) scheme.
The computational complexity of the LDPC+CBI+Polar(SC) system is
lower than  LDPC+Polar(BP) systems at a relatively small cost of the delay.

\section{Conclusion}\label{sec_con}
In this paper, we proposed a novel interleaving scheme, the correlation-break interleaving (CBI),
which can improve the BER performance of polar codes
while still maintaining the low complexity of the SC decoding of polar codes.
The CBI scheme has a small average delay compared with a blind interleaving scheme.
Simulation results are provided which verified that the concatenation of polar codes
with SC decoding
and the CBI scheme achieves almost the same BER performance as the concatenation scheme
of polar codes with the BP decoding.
\bibliography{../ref_polar}
\bibliographystyle{IEEEtran}
\end{document}